\documentclass[aps,pra,preprint,floats]{revtex4}

\usepackage{graphicx}
\usepackage{amsmath}
\newcommand{\beq}{\begin{equation}}
\newcommand{\eeq}{\end{equation}}
\newcommand {\ket}[1]{|\,{#1}\,\rangle}
\newcommand {\bra}[1]{\langle\,{#1}\,|}
\begin{document}

\title{Molecular Response in One Photon Absorption: Pulsed
Laser Excitation vs. Natural Thermal Light}
\author{Paul Brumer$^*$}
\affiliation{Chemical Physics Theory Group, Department of Chemistry,
University of Toronto, Toronto, Ontario, Canada M5S 3H6}
\author{Moshe Shapiro}
\affiliation{Department of Chemistry, University of British Columbia,
Vancouver, British Columbia, Canada,  V6T 1Z3}

\maketitle

\vspace{0.5in}

\noindent

CLASSIFICATION: Physical Sciences, Chemistry

\noindent
$^*$ Corresponding Author: Paul Brumer, Department of Chemistry, University of
Toronto, Toronto, Ontario, Canada M5S 3H6, Tel: 416-978-3569,
email:pbrumer@chem.utoronto.ca

\newpage

{\bf Abstract}

Photoinduced biological processes occur via one photon absorption
in natural light, which is weak, CW and incoherent, but are often studied in the laboratory
using pulsed coherent light. Here we compare the response of a
molecule to these two very different sources within a quantized radiation
field picture. The latter  is
shown to induce coherent time evolution in the molecule,
whereas the former  does not. As a result, the coherent time dependence observed
in the laboratory experiments will not be relevant to the natural
biological process.
Emphasis is placed on resolving 
confusions regarding
this issue that are shown to arise from
aspects of quantum measurement and from a lack of appreciation of  the proper
description of the absorbed photon.

\section{Introduction}

The nature of the molecular response to weak electromagnetic
fields, where the probability of
absorbing a photon is small, is the subject of considerable
importance in light-induced biological processes.  Examples include
light harvesting complexes\cite{flemingreview,scholeslessons,rienk}
and vision\cite{earlyretinal,earlyretinal2,miller,hoki,newretinal}, both of which operate in the domain of weak
photon flux.

Recent experimental studies have generated
considerable excitement\cite{wolynes,ball,lloyd} due to the observation of
long lived coherent (electronic and vibrational \cite{turner}) quantum time evolution 
subsequent to pulsed laser excitation of various 
biomolecules\cite{fleming,fleming2,scholes,engel1,engel2,leonardo,coker}.
Similar enthusiasm\cite{martinez} has been generated by the coherent vibrational
dynamics observed in retinal isomerization induced by pulsed laser
light\cite{earlyretinal,newretinal}. These references, as well as 
many others, either explicitly or implicitly assume that the observed 
coherent time evolution is of considerable biological significance.

Of particular relevance, then, is whether the observed coherent time
evolution does, indeed, play a biological role. That is, is
the molecular response in laboratory laser experiments
that use pulsed coherent laser light\cite{fleming,scholes,miller,engel1}
relevant when the system is irradiated with natural
light, i.e. radiation arising from a thermal source
that is essentially CW and highly
incoherent\cite{xupei,hoki,mancal}?
This issue (albeit not biologically motivated)
was treated some time ago using a semiclassical approach to the
light-matter interaction within first-order perturbation theory\cite{xupei},
leading to the conclusion that the responses are very different:
isolated molecules subject to
pulsed coherent laser light display subsequent coherent time evolution, whereas those
subject to incoherent light from a thermal CW source do not.
In addition, that study showed that pulsed incoherent light, which
by definition is partially coherent,
induces time evolution on the time scale of the pulse.
i.e. the molecule responds to the time envelope of the light pulse.
However for sunlight, for example, the time scale of the envelope  is
hours, and a stationary non-evolving state is reached almost immediately.

These results have profound implications for
biological processes induced by weak fields (photosynthesis, vision),
where the probability of single photon absorption is small due to the
low photon flux.  They have, however, been largely ignored, and have 
recently  been
confused by arguments based on a qualitative picture of photons and of
photon-molecule interactions.

For example, a
current qualitative description\cite{flemingbad} suggests that the absorption
of a single photon triggers the same coherent
molecular response,  regardless of
the character  of the light source. There the view is
that a single photon, incident
on a molecule, whether arising from a pulsed coherent laser source or from a
natural incoherent CW blackbody source (such as the sun), ``kicks" the molecule
and undergoes coherent time evolution.
Further, there are
related concerns within the community
associated with times of arrival of the photons, the
role of different bases that can be used to describe the incident light,
etc. Clearly, clarifying these issues is necessary to understanding coherent quantum 
processes in biology, and
benefits from a proper quantized picture
of the photon and its interaction with molecules, utilized in this paper.

This paper, which has two parts, addresses these issues. In the first part
we formulate the problem of one-photon absorption using
quantized radiation fields (extensive introductions to the approach used
below are provided in Refs. \cite{loudon,ourbook1,ourbook2}).
This quantized radiation field approach
provides a focus on the role of the photon,
explicitly displays issues
related to light-matter entanglement, permits consideration of an
expanded collection of photon sources, and allows us to obtain results without
requiring details of light-molecule time evolution.
This treatment clearly shows
that one photon absorption from a pulsed coherent source induces coherent molecular
dynamics whereas one photon absorption from a natural incoherent CW thermal source
does not. In the second part, we utilize these results to
provide qualitative insight into
the nature of the photon and its role in comparing pulsed coherent
laser excitation to excitation with natural light.
Specifically, we emphasize that
the description of the photon necessarily carries with it information
about the source of the radiation, and that problems that have arisen in
qualitatively understanding this process and its role in biology result from
(a) a simplified
view of the particulate nature of the incident light and of light-matter
interactions, and (b) an incomplete understanding of issues in quantum
measurement theory.

Three initial clarifying remarks are in order:

(a) The literature, in discussing the role of
``quantum coherence" in biological processes uses the term in two different ways.
The first, relevant here\cite{earlyretinal,earlyretinal2,newretinal,fleming,fleming2,scholes,engel1},
refers to coherent time evolution of a system that is, by definition, associated
with coherent superpositions of nondegenerate eigenstates of the Hamiltonian.
That is, off-diagonal elements of the system density matrix in the energy representation 
$\rho_{jk}$ evolve with phases of the form $\exp[i(E_j-E_k)t/\hbar]$, where the $E_i$ are
energy eigenvalues of the system. 
The second, unrelated to the issue addressed here, refers to the character of
the stationary energy eigenstates
that span numerous subcomponents within the overall system (e.g. various
molecular sites within a photosynthetic complex\cite{schulten}).

(b) We emphasize that considerations below apply to an isolated system.
Open systems coupled to an environment are discussed
elsewhere using a semiclassical approach to light-matter 
interactions\cite{hoki,mancal} with similar qualitative conclusions. 
In addition, open quantum system and related scenarios 
introduce yet a third 
definition of the word ``coherences". Specifically, open quantum systems 
permit the existence of off-diagonal system density matrix elements $\rho_{jk}$
which do not evolve in time, or which appear in steady state scenarios.
These are ``stationary coherences" (see e.g.
Ref. \cite{mancal,scully1} or the time independent $\rho_{12}$ in Eq. (14)
of Ref. \cite{scully2}) which are distinctly different from the time evolving coherences
which are the focus of this paper.

(c)  Changes in the populations of energy eigenstates of the system, without the
involvement of time-dependent off-diagonal $\rho_{jk}$ are also mentioned below,
where they are termed ``incoherent dynamics".

\section{One-Photon Absorption}

Consider the interaction
of radiation with an isolated material system that is
initially in a stationary state $\ket{E_i}$.
For notational convenience
this state is assumed non-energetically degenerate.
Higher eigenstates of energy $E_j$ are denoted $\ket{E_j,{\bf m}}$, where
${\bf m}$ denotes any additional quantum numbers needed to describe the
state. States of the radiation field are described below in terms of
number states $\ket{N_k}$.
Here $N_k$ is the
number of photons in the $k^{\rm th}$ mode, of frequency $\omega_k$, and
$k=1,...N$ is a (plane wave) mode index.

\subsection{Coherent Sources}

Consider now absorption from an arbitrary radiation field.
A general field  of this kind,
linearly polarized along the $\hat{{\bf \epsilon}}$ direction,
can be parametrized as a sum of superpositions
of products of number states $\ket{N_i}$:
\beq
\ket{R_i}=
\hat{{\bf \epsilon}}\sum_{N_1,N_2,...,N_{max}}
c(N_1,N_2,...,N_{max})
\ket{N_1}\ket{N_2}...\ket{N_{max}},
\eeq
For computational simplicity we also use the notation
\beq
\ket{R_i} = \hat{{\bf \epsilon}}\sum_{\bf N} c({\bf N})
\ket{{\bf N}}.
\label{initstate}
\eeq
where ${\bf N} =
(N_1,N_2,...N_{max})$.

For example, output from a standard multi-mode laser source can be
well represented as a product
$\ket{R_i} = \prod_k \ket{\alpha_k}$
of coherent states $\ket{\alpha_k}$, where\cite{loudon,chaio}
\beq
\ket{\alpha_k} = \exp(-|\alpha_k|^2/2) \sum_{N_k} \frac{\alpha_k^{N_k}}{(N_k!)^{1/2}}\ket{N_k},
\label{costate}
\eeq
i.e.,
\beq
c(N_1,N_2,...,N_{max}) = \prod_{k=1}^{N_{max}}
\exp(-|\alpha_k|^2/2)  \frac{\alpha_k^{N_k}}{(N_k!)^{1/2}}.
\label{cohstates}
\eeq
The larger the parameter $\alpha_k$, the closer the radiation is to classical
light.

Consider then the interaction
of the radiation field  with an isolated material system that is
initially in a stationary state $\ket{E_i}$.
The initial radiation-matter state is then given by
\beq
\ket{\Psi_i}=\ket{R_i}\ket{E_i}.
\eeq
Assuming the dipole approximation and using first order perturbation theory,
the final state,
after absorbing {\it one photon} from the field, becomes a
radiation-matter wave packet, in the excited state, of the form
\beq
\ket{\Psi_f}=
\sum_{k,{\bf m},{\bf N}} |A(k,{\bf m})\rangle
 c({\bf N})
\ket{N_1}...\ket{N_{k-1}}\ket{N_k-1}\ket{N_{k+1}}...\ket{N_{max}},
\label{finalwf}
\eeq
where							
\beq
|A(k,{\bf m})\rangle =
\frac{2\pi i}{\hbar}\varepsilon(N_k,\omega_k)
\ket{E_k,{\bf m}}\bra{E_k,{\bf m}}\hat{{\bf \epsilon}}\cdot {\bf d}\ket{E_i}.
\eeq
Here ${\bf d}$ is the electric dipole operator and
$E_k=E_i+\hbar\omega_k$ is the energy imparted to the
material system as a result of the
absorption of one photon of frequency $\omega_k$.
The field amplitude $\varepsilon(N_k,\omega_k)$ introduced above is defined as
\beq
\varepsilon(N_k,\omega_k)= i
\left(\frac{\hbar\omega_k N_k}{\epsilon_0 V}\right)^\frac{1}{2} \exp(i\omega_kz/c),
\label{varepsilon} \eeq
with $z$ denoting the
axis of propagation of the light beam, $\epsilon_0$ is the
permittivity of the vacuum, and $V$ is the cavity volume. Note that the
resultant state [Eq. (\ref{finalwf})] is an entangled superposition of the
states of the molecule and the radiation field\cite{bsfaraday}.

The density matrix $\rho_f$ associated with $\ket{\Psi_f}$ is given by
$$
\rho_f = \ket{\Psi_f}\bra{\Psi_f} =
\sum_{k,k',{\bf m},{\bf m}',{\bf N},{\bf N}'} |A(k,{\bf m})\rangle
\langle A(k',{\bf m}')| c({\bf N})c^*({\bf N}') \times $$
\beq
\ket{N_1}...\ket{N_{k-1}}\ket{N_k-1}\ket{N_{k+1}}...\ket{N_{max}}
\bra{N^{\prime}_1}...\bra{N^{\prime}_{k'-1}}\bra{N^{\prime}_{k'}-1}
\bra{N^{\prime}_{k'+1}}...\bra{N^{\prime}_{max}},
\label{totaldens}
\eeq

Our interest lies in the state of the system, as opposed to the state of
the (system + radiation field). In accord with standard
quantum mechanics\cite{schlosshauer}, one extracts this information from $\ket{\Psi_f}$ by
constructing the density matrix $\rho_f=\ket{\Psi_f}\bra{\Psi_f}$ and tracing over
the radiation field to give the density matrix of the molecule, denoted
$\rho_{mol}$. Doing so, gives:

\begin{eqnarray}
&&\rho_{mol} = \sum_{{\bf N}''} \langle{\bf N}''| \rho_f \ket{{\bf N}''} \nonumber \\
&=& \sum_{{\bf N},{\bf m}, {\bf m}',k} |c({\bf N})|^2
|A(k,{\bf m})\rangle \langle A(k,{\bf m}')| +
\sum_{{\bf N},{\bf m}, {\bf m}', k'>k} [~d_{k',k} | A(k,{\bf m})\rangle
\langle A(k',{\bf m}')| + cc~]
\label{densmatrix}
\end{eqnarray}
where $cc$ denotes the complex conjugate of the term that precedes it, and
\beq
d_{k',k} = c(N_1,N_2,..,N_{k' -1}, N_{k'} -1, N_{k'+1}...,N_{max})~~
c^*(N_1,N_2,..,N_{k -1}, N_k -1, N_{k+1}...,N_{max}).
\label{dmatrix}
\eeq

Consider now a coherent pulse of light.
If $t_0$ denotes the time at which the pulse is over then, given
Eq. (\ref{densmatrix}), the molecule will evolve after $t_0$ as

\begin{eqnarray}
\rho_{mol}(t > t_0) &=&
\sum_{{\bf N},{\bf m}, {\bf m}',k} |c({\bf N})|^2
|A(k,{\bf m})\rangle \langle A(k,{\bf m}')|  \nonumber \\
&+&2\sum_{{\bf N},{\bf m}, {\bf m'}, k'>k} {\rm Re} [~d_{k',k}
|A(k,{\bf m})\rangle \langle A(k',{\bf m}')|
\exp[-i(E_k-E_{k'})(t-t_0)/\hbar ]]
\label{densmatrixevolve}
\end{eqnarray}
For example, for coherent states the real, positive $d_{k,k'}$ can be composed
from Eq. (\ref{cohstates}).

It is clear from Eq. (\ref{densmatrixevolve})
that one photon  absorption from the coherent
pulse of light excites
many material states, producing a {\it coherent molecular
superposition state} that evolves coherently in time with frequencies
$(E_k-E_{k'})/\hbar$.
The energy of this
superposition state, which is composed
of many $\ket{E_k,{\bf m}}$ eigenstates, is not sharply
defined. The fact that the system
evolves coherently in time after pulsed coherent light absorption is intimately tied to this
energy uncertainty. This, in turn, arises from the
fact that $\ket{R_i}$ is itself a superposition of
non-energetically degenerate states of the radiation
field.

\subsection{Natural Incoherent Thermal Sources}

Consider now absorption of a photon that is emitted by an
incoherent thermal source, such as sunlight.
This source consists of
a statistical \textit{mixture} of number states described by a radiation field
density matrix\cite{loudon}:
\beq
\rho_R = \sum_{\bf N} p_{\bf N}|{\bf N}\rangle\bra{{\bf N}}.
\label{mixedrad}
\eeq
Here $p_{{\bf N}}$ is the probability of finding the number state
$\ket{{\bf N}}$ in the radiation emitted from the thermal source.
If the source is at temperature $T$ this is given by:
\beq
p_{{\bf N}} = \prod_k \frac{(\overline{N}_k)^{N_k}}{(1+\overline{N}_k)^{1+N_k}}
\eeq
with $\overline{N}_k$ being the mean number of photons at temperature $T$:
$\overline{N}_k = [ \exp(\hbar \omega_k /k_BT)-1]]^{-1}$.

This  radiation field is a statistical mixture of
number states. As a consequences, irradiation with this source will yield
an uncorrelated mixture of states resulting from
excitation with the state $|{\bf N}\rangle\langle{\bf N}|$.
Excitation with the single state $|{\bf N}\rangle\langle{\bf N}|$ can be obtained from
the above treatment by setting $$c({\bf N})c^*({\bf N'})=|c({\bf N})|^2 \delta_{{\bf N}.{\bf N'}}$$
in Eq. (\ref{totaldens}).
In this case, in Eq. (\ref{dmatrix}) $d_{k,k'} = \delta_{k,k'}$ and Eq. (\ref{densmatrix})
becomes
\beq
\rho_{mol}
=\sum_{{\bf m}, {\bf m}'} |A(k,{\bf m})\rangle \langle A(k,{\bf m}')|
\label{incohdens}
\eeq
Hence, the result of one-photon excitation with radiation emitted by a thermal
incoherent CW source [Eq. (\ref{mixedrad})] would be given by an incoherent
weighted sum over Eq. (\ref{incohdens}).

The system, after one-photon excitation, is then in a mixture of stationary states,  and $\rho_{mol}$ does not
subsequently evolve coherently in time. Rather, as the natural light continues to stay on for times 
long compared
to molecular time scales, the subsequent time evolution is entirely incoherent,
with the populations of the energy eigenstates evolving incoherently in accord with
Einstein's rate laws\cite{zoller}.

\section{Discussion and Summary}

The results of the above analysis are clear, but a discussion is
warranted. Absorption of one photon from a
coherent pulse creates a superposition of energy eigenstates, and hence
a state that evolves coherently in time.
By contrast, absorption from a thermal incoherent CW source such as the sun
is seen to create a stationary mixture.
The qualitative results of this quantized-radiation field analysis  of one
photon absorption is in agreement with that  obtained\cite{xupei}
in a treatment using semiclassical light-matter interactions.
What is clearer here, however, is the specific
focus on the absorption of a single photon. This analysis can now be used
to comment on the associated physics and on current concerns
that have arisen regarding  one-photon absorption.

Recent qualitative considerations
have led to incorrect conclusions, such as that the coherence of the
molecule, post-excitation,  is independent of the nature of the radiation source\cite{flemingbad}.
Related incorrect pictures
have also arisen, suggesting, for example, that each incident photon in weak
CW light gives the molecule ``a kick", which induces dynamics in the
molecule.
These views,  not supported by the above analysis, arise from a
simplistic particle picture of the photon\cite{rajroy}, and are dispelled when one
appreciates the role of measurement in quantum mechanical particle/wave
duality. That is,
as is typically the case, whether a system behaves like a wave or a particle
depends upon the nature of the measurement\cite{steinberg}. For example,
in the case of pulsed light absorption described above,
no measurement is made that would reveal particle-like properties of the
photon. Hence, utilizing language associated with
a particle picture is not correct for this physical
scenario.

Analogously, for the pulse case,
were one to undertake an experiment in which measurements of  the
energy of the molecule subsequent to absorption of light from a pulse were made,
then stationary states at fixed energy would emerge. Such a measurement is not
done, and hence the energy of the system is uncertain, which is
intimately related to the fact that the molecule undergoes coherent 
time evolution. By contrast, a thermal source,
by its very nature, is comprised of independent fixed energy photons and,
as such, creates stationary molecular states upon irradiation.
That is, conservation of energy
ensures that an initial energy eigenstate, absorbing a single photon of known
energy, reaches a stationary excited state with known energy, and no 
subsequent coherent time evolution.

Similarly, adopting a classical picture of the photon as a particle incident on the
molecule, possessing only information about its energy and polarization, and
possessing no characteristics associated with the
source of the radiation, is incorrect. Specifically, as is evident from the
analysis above, the effect that the photon has on the molecule depends intimately
on the nature of the light source. Multimode coherent pulses induce coherent
dynamics whereas CW sources (and likewise natural thermal sources) do not.

The classical picture of the photon as a particle incident on the molecule,
repeatedly initiating dynamics, also assumes a known photon arrival time. This
too is incorrect and inconsistent with the quantum analysis insofar as no
specific arrival time can be presumed unless the experiment itself is
designed to measure such times.

Finally, suggestions  have been made that a thermal source may
be regarded as a collection of random femtosecond
pulses.  As such, the suggestion goes,
each molecule feels the effect of individual femtosecond pulses
and undergoes coherent time-dependent evolution\cite{flemingreview}. 
this perspective
is also unjustified. Specifically,
there is no justification for imposing a specific
physical picture associated with
femtosecond pulses on the natural process
if the natural scenario makes no such measurement.
That is, the electric field from a
thermal light source can be expanded in a variety of different
bases. However, (a) at best the expansion should be done in a basis
related to the physics, i.e. a source of spontaneous emission that is
phase interrupted\cite{loudon}, and (b) regardless of the basis used,
it is the {\it overall} effect of the light that is important,
and this overall effect is to populate energy eigenstates of the molecule, 

One final note is in order. As is well-known, even thermal sources will create {\it very short time} initial
coherences associated with the initial time that the molecule feels the
turn-on of the light. Short time coherent dynamics is then manifest.
However, under natural circumstances (such as moonlight or sunlight)
such initial time evolution (on the order of tens of fs) 
is totally irrelevant on the time scales
associated with natural light\cite{hoki}. That is, after this short time,
the system is in a mixture of stationary states.

It is worthwhile, nonetheless,
to appreciate the character of such initial dynamical coherences.
Consider, for example,
natural thermal light incident on Pyrazine.
Here the well-known excitation is from the S$_0$ electronic ground state to
an S$_2$ excited state\cite{pyrazine1,pyrazine2,christopher}. This S$_2$ state is, in turn, coupled to an S$_1$ state,
which will be occupied as
the CW light drives the system into stationary states. Hence, on the short
time scale there is coherent S$_2$ to
S$_1$ internal conversion, since (a) the S$_2$ state is the bright state
that is created upon excitation, and
(b) the exact energy eigenstates to which the system is driven
by the CW light contains density on both the S$_2$ and S$_1$ electronic
states. Hence there is initial coherent transient dynamics.
However, this coherent dynamics does not continue after the
short transient time. Rather, in accord with the analysis above, since the light is
thermal, no coherent
molecular dynamics will occur after the brief initial transient. Rather, the
population of the stationary states will change incoherently 
without the establishment
of coherence between energy levels, and the ratio of the population of
S$_2$ to S$_1$ will be unchanged as time progresses.

In summary, one photon molecular excitation and with 
pulsed coherent laser light and natural incoherent light yield qualitatively
different responses. Further,
the above approach makes clear the extent to which quantum
mechanics allows a physical picture of one photon absorption in an isolated
molecule. An analogous picture arises in open systems\cite{hoki,mancal}.
Any additional imposed qualitative picture may well be inconsistent with
quantum mechanics.

{\bf Acknowledgments}:
Financial support from the U.S. Air Force Office of
Scientific Research under grant number FA9550-10-1-0260 and from the
NSERC, Canada  is gratefully acknowledged.

\end{document}